\documentclass[11pt]{article}
\usepackage{amsmath}
\usepackage{amsfonts}
\usepackage[dvips]{graphicx}
\usepackage[hang,centerlast,small,sc]{caption}
\usepackage{xspace}
\usepackage[mathscr]{eucal}

\newcommand{\gama}[1]{\mbox{$\Gamma(#1)$}}
\newcommand{\kmax}{\ensuremath{k_{\rm max}}\xspace}
\newcommand{\kbar}{\ensuremath{\overline{k}}\xspace}
\newcommand{\calGk}{\ensuremath{\mathcal G_{k}}\xspace}

\begin{document}
\title{\bf 
Bayesian finite mixtures: a note on prior specification and posterior computation}
\author{Agostino Nobile\thanks
{Department of Statistics, University of Glasgow, Glasgow G12 8QW, Scotland.
\newline \hspace*{5mm} Email: {\tt agostino@stats.gla.ac.uk}}\\
University of Glasgow, Scotland.}

\date{May 2005}
\maketitle
\begin{center}
{\large \bf Abstract}

\end{center}
\hspace*{\fill}

\parbox{6in}{
{\small
A new method for the computation of the posterior distribution of the number $k$ of components in a finite mixture is presented. Two aspects of prior specification are also studied: an argument is made for the use of a $Poi(1)$ distribution as the prior for $k$; and methods are given for the selection of hyperparameter values in the mixture of normals model, with natural conjugate priors on the components parameters.
}
}
\hspace*{\fill} \\[\bigskipamount]


\pagenumbering{arabic}

\section{Introduction}
\label{Sec-intro}

Finite mixture distributions have become widely used as a tool of semi-parametric inference: they partake of the conceptual simplicity of parametric models and of the flexibility of non-parametric ones. 
This paper is a contribution to the Bayesian analysis of finite mixtures with an unspecified number of components. 
I give arguments to support the use of a $Poi(1)$ prior for the number of components and present a new method for the numerical computation of its posterior. The method exploits a fundamental probability identity already used by Chib~(1995), but combines it with the representation of mixture marginal likelihoods given in Nobile~(2004). I also discuss a more specific topic, hyperparameter selection in a finite mixture of univariate normals. Throughout the paper, the galaxy data set is used for illustrative purposes. 

The remainder of this section provides a brief introduction to Bayesian finite mixtures, representations of the associated marginal likelihoods and the mixture of normals model. Section~\ref{Sec-marlik} deals with the estimation of marginal likelihoods using the frequency of empty components in a Markov Chain Monte Carlo sample of the mixture allocations. Section~\ref{Sec-priork} argues that the structure of the model suggests the $Poi(1)$ distribution as a suitable prior for the number of components, when no substantive information on it is available. Section~\ref{Sec-comparam} concerns the more practical issue of hyperparameter determination in the mixture of normals model, when natural conjugate priors on the means and variances of the components are employed.

\subsection{Bayesian finite mixtures}

A finite mixture is a distribution with density, with respect to some underlying measure, given by
\begin{equation}
	f(x) = \sum_{j=1}^{k} \lambda_{j} p_{j}(x |\theta_{j}).
							\label{finmix}
\end{equation}
The weights $\lambda_{j}$ are non-negative and sum to 1, while the component densities $p_{j}(\cdot|\theta_{j})$ belong to some known parametric family.
Observations $x_{1}, \ldots, x_{n}$ are regarded as proceeding from the distribution (\ref{finmix}) and interest lies in the number of components $k$ and, conditional on $k$, in the weights $\lambda$ and the components' parameters $\theta$. 

The model can be rewritten by introducing latent allocation vectors $g=(g_{1}, \ldots, g_{n})$ with $g_{i} \in \{1,\ldots,k\}$ denoting the mixture component that generated the $i$-th observation:

\parbox{14cm}{
	\begin{eqnarray*}
	\Pr[g_{i} = j | k, \lambda] & = & \lambda_{j} 
		\hspace{60pt} i=1,\ldots,n,
		\hspace{10pt} \mbox{independently} \\ 
	x_{i} | k, g, \lambda, \theta & \stackrel{\mathrm{ind.}}{\sim} & 
		p_{j}(x_{i} | \theta_{j}),
		 \hspace{25pt} j=g_{i}, \hspace{10pt} i=1,\ldots,n.
	\end{eqnarray*} 
} \hfill
\parbox{1cm}{\begin{eqnarray}\label{finmixg}\end{eqnarray}}

\noindent
In the Bayesian analysis of the model, typically one assumes that $\lambda | k \sim Dir(\alpha_{1}, \ldots, \alpha_{k})$, where the $\alpha_{j}$'s are fixed constants.
Also, the component parameters $\theta_{j}$ are assumed a priori independent, conditionally on $k$ and, possibly, a vector of hyperparameters $\phi$:
\[
	\pi(\theta|k,\phi) = \prod_{j=1}^{k} \pi_{j}(\theta_{j}|\phi_{j}).
\]

If a prior distribution $\pi(k)$ is specified, then one can obtain a sample from the joint posterior of $(k, \lambda, \theta)$ by means of Markov chain Monte Carlo methods, see e.g. Richardson and Green~(1997), Phillips and Smith~(1996), Stephens~(2000a), Nobile and Fearnside~(2005). 
Inference about $\lambda$ and $\theta$ is not straightforward, because the likelihood is invariant with respect to permutations of the components' labels. Achieving identifiability by imposing constraints on the parameters does not always work and other methods have been proposed, see Richardson and Green~(1997) and its discussion (especially the contributions of G. Celeux and M. Stephens), Celeux, Hurn and Robert~(2000), Stephens~(2000b), Fr\"{u}hwirth-Schnatter~(2001), Nobile and Fearnside~(2005).

An alternative to sampling from the posterior of $(k, \lambda, \theta)$ consists of estimating the marginal likelihoods $f_{k}$ of the mixture model with $k$ components:
\[
	f_{k} := f(x|k) = \iint f(x|k, \lambda, \theta)
				\pi(\lambda, \theta | k)
				\, \mathrm{d} \lambda \, \mathrm{d} \theta,
		\hspace{30pt} k=1,2,\ldots,\kmax.
\]
Each marginal likelihood estimate makes use of MCMC output for a model with fixed $k$. The estimates can be used to compute Bayes factors for $k$ vs. $k-1$ components, or to compute the posterior of $k$, $\pi(k|x) \propto \pi(k) f_{k}$. It should be noted that estimation of the marginal likelihood from MCMC output is not as simple as other posterior inference using MCMC, and as a consequence several methods have been proposed, see e.g. Chib~(1995), Raftery~(1996), DiCiccio et al. (1997), Gelman and Meng~(1998) and references therein.

\subsection{Marginal likelihoods of finite mixtures}
\label{Sec-intro-marlik}

The marginal likelihoods can be rewritten as 
\begin{equation}
	f_{k} = \sum_{g \in \calGk} f(g|k) f(x|k,g) 
							\label{fk}
\end{equation}
where the sum extends over the set \calGk of all the allocation vectors with entries less than or equal to $k$, see e.g. Nobile~(1994, 2004). In equation (\ref{fk}), $f(g|k) = \int f(g|k,\lambda) \pi(\lambda|k) \, \mathrm{d} \lambda = \frac{\gama{\alpha_{0k}}}{\gama{\alpha_{0k}+n}} {\displaystyle \prod_{j=1}^{k}} \frac{\gama{\alpha_{j}+n_{j}}}{\gama{\alpha_{j}}}$ with $\alpha_{0k}=\sum_{j=1}^{k} \alpha_{j}$ and $n_{j}$ equal to the number of observations that $g$ allocates to component $j$. 
The other term in the right hand side of (\ref{fk}), $f(x|k,g)$, is obtained by integrating $f(x|k,g,\theta)$ from (\ref{finmixg}) with respect to the prior distribution of $\theta|k$.
Although this integration can be performed in closed form only for some prior distributions, notably natural conjugate priors on $\theta$, representation (\ref{fk}) is always valid.  
Under the assumption that the Dirichlet hyperparameters $\alpha_{j}$ and the prior distributions $\pi_{j}(\cdot|\phi_{j})$ remain the same for fixed $j$ as $k$ varies, 
the marginal likelihoods enjoy further representations. Partition the set of allocation vectors \calGk as 
\[
	\calGk = \bigcup_{t=1}^{k} \mathcal{G}_{t}^{\star},
	\hspace{30pt}
	\mathcal{G}_{t}^{\star} \cap \mathcal{G}_{s}^{\star} = \emptyset,
	\hspace{20pt} t \neq s
\]
where $\mathcal{G}_{t}^{\star}$ is the set of allocation vectors which assign at least one observation to component $t$ and none to higher components. Also, let $f^{\star}_{t}$ be the portion of the marginal likelihood $f_{t}$ that accounts for vectors $g$ allocating at least one observation to component $t$ and none to higher: 
\begin{equation}
	f^{\star}_{t} = \sum_{g \in \mathcal{G}_{t}^{\star}} 
				f(g|t) f(x|t,g).
							\label{fstart}
\end{equation}
Then one can show (see Nobile 2004, page 2049) that, for all $g \in \mathcal{G}_{t}^{\star}$ with $t < k$,
\begin{equation}
	f(x|k,g)  =  f(x|t,g)
	\hspace{30pt} \mbox{and} \hspace{30pt}
	\frac{f(g|k)}{f(g|t)}  = 
		\frac{\gama{\alpha_{0k}}}{\gama{\alpha_{0k}+n}}
		 \frac{\gama{\alpha_{0t}+n}}{\gama{\alpha_{0t}}}
	=: a_{kt}
							\label{C1C2}
\end{equation}
and that 
\begin{eqnarray}
	f_{k} & = & \sum_{t=1}^{k} a_{kt} f^{\star}_{t} \label{fk1} \\
		& = & a_{k,k-1} f_{k-1} + f^{\star}_{k} \label{fk2}.
\end{eqnarray}

If the prior distribution of $(\lambda, \theta)|k$ is invariant to permutations of the components labels, a stronger result is available. Let $\mathcal{G}^{t}_{h}$ be the subset of $\mathcal{G}^{\star}_{t}$ consisting of allocations with $h \leq t$ non-empty components. 
In particular, any vector $g \in \mathcal{G}^{h}_{h} \subset \mathcal{G}_{h}$ assigns at least one observation to each mixture component $1, \ldots, h$.
Let $f_{h}^{\dagger}$ be the portion of $f_{h}$ which corresponds to allocations with no empty components:
\[
	f_{h}^{\dagger} = \sum_{g \in \mathcal{G}_{h}^{h}} 
				f(g|h) f(x|h,g). 
\] 
Then (Nobile 2004, page 2053)
\begin{equation}
	f_{k}  =  \sum_{h=1}^{k \wedge n} 
			\binom{k}{h} a_{kh} f_{h}^{\dagger}. \label{fk3}
\end{equation}
For related representations see Ishwaran, James and Sun~(2001).

\subsection{Mixtures of univariate normals}
\label{Sec-univnorm}
The method to be presented in the following section is of general applicability. Since mixtures of univariate normals will be used as an illustration, I introduce here some notation.
It is assumed that the component densities $p_{j}(\cdot|\theta_{j})$ are normal with mean $m_{j}$ and variance $r^{-1}_{j}$:
\[
	x_{i} | k, g, \lambda, \theta  
		\;\; \stackrel{\mathrm ind.}{\sim} \;\; 
		N(m_{j}, r^{-1}_{j}),
		 \hspace{25pt} j=g_{i}, \hspace{10pt} i=1,\ldots,n.
\]
Independent natural conjugate priors are placed on $\theta_{j}=(m_{j},r_{j})$, $j=1,\ldots,k$:

\parbox{14cm}{
	\begin{eqnarray*}
		r_{j} & \stackrel{\mathrm{ind.}}{\sim} & Ga(\gamma,\delta) \\
		m_{j} |r_{j}   & \stackrel{\mathrm{ind.}}{\sim} & 
				N(\mu, \{\tau r_{j} \}^{-1} ),
	\end{eqnarray*}
} \hfill
\parbox{1cm}{\begin{eqnarray}\label{normprior}\end{eqnarray}}

\noindent
with $E(r_{j}) = \gamma / \delta$.
See Diebolt and Robert~(1994) or Nobile and Fearnside~(2005) for more details.
Other priors on $\theta$, such as the independent prior used by Richardson and Green~(1997), could be used as well.

The prior distribution (\ref{normprior}) requires the specification of four hyperparameters. The overall mean $\mu$ is set to a round value close to the sample mean $\overline{x}$. The shape parameter $\gamma$ is half the degrees of freedom of the prior predictive $t$ distribution. I choose $\gamma=2$, to have a $t_{4}$ prior predictive, with relatively thick tails, but finite second order moments. The choice of the scale parameter $\delta$ and of $\tau$, the prior ratio between within components variance and variance of the means, is discussed in Section~\ref{Sec-comparam}.

\section{Marginal likelihoods from empty components}
\label{Sec-marlik}

For a given parametric model $f(x|\theta)$ and prior distribution $f(\theta)$, the marginal likelihood of the observed data $x$ is defined as $f(x) = \int f(x|\theta) f(\theta) \, \mathrm{d} \theta$. Using Bayes theorem, $f(x)$ can be rewritten as 
\begin{equation}
	f(x) = \frac{f(\theta) f(x|\theta)} {f(\theta|x)}
							\label{marlikgen} 
\end{equation}
where $f(\theta)$ and $f(x|\theta)$ are assumed computable, including their normalizing constants, and the formula holds for any parameter value $\theta$. 
Expression (\ref{marlikgen}) forms the basis of a method of marginal likelihood estimation, see Chib~(1995) and Raftery~(1996). In short, although typically the posterior $f(\theta|x)$ cannot be evaluated exactly, an estimate of it at some parameter value $\theta$ can be obtained using a Monte Carlo sample; substituting this estimate into (\ref{marlikgen}) yields an estimate of $f(x)$.

In the context of Section~\ref{Sec-intro-marlik}, the marginal likelihood for the model with $k$ components can be written as
\begin{equation}
	f_{k} = \frac{f(g|k) f(x|k,g)}{f(g|k,x)}.
							\label{marlik}
\end{equation}
Here the allocation vector $g$ plays the role of $\theta$ in the above discussion and everything is conditional on $k$.
Since (\ref{marlik}) holds for all $g \in \calGk$, it still holds if one sums both numerator and denominator over any non-empty set $E$:
\begin{equation}
	f_{k} = \frac{\sum_{g \in E} f(g|k) f(x|k,g)}
				{\sum_{g \in E} f(g|k,x)}.
							\label{marlikE}
\end{equation}
Letting $E=\mathcal{G}^{\star}_{k}$, the denominator of (\ref{marlikE}) is the posterior probability of $\mathcal{G}^{\star}_{k}$, while the numerator equals $f^{\star}_{k} = f_{k} - a_{k,k-1} f_{k-1}$, using equations (\ref{fstart}) and (\ref{fk2}). One then obtains 
\[
	f_{k} = \frac{f_{k} - a_{k,k-1} f_{k-1}}
			{\Pr[\mathcal{G}^{\star}_{k} |k, x]}
\]
and after rearranging
\begin{equation}
\frac{f_{k}}{f_{k-1}} = 
		\frac{a_{k,k-1}} {1 - \Pr[\mathcal{G}^{\star}_{k} |k, x]}.
							\label{BF}
\end{equation}
The left hand side of (\ref{BF}) is the Bayes factor $B_{k,k-1}$ for the model with $k$ components against the model with $k-1$ components. In the right hand side $a_{k,k-1}$ is a known constant, while the denominator is the posterior probability, according to the model with $k$ components, that the $k$-th component is empty, which can be easily estimated using a MCMC sample from $f(g|k,x)$. In some mixture models $f_{1} = f^{\star}_{1}$ is computable exactly; if this is the case, estimates of the marginal likelihoods, if needed, can be readily produced from the Bayes factors. Otherwise, one can still obtain estimates of normalized marginal likelihoods, by setting $f_{1}=1$ and then rescaling the sequence of $f_{k}$'s. 

Using formula (\ref{BF}) is somewhat wasteful, since it only employs the fixed $k$ MCMC sample to estimate $\Pr[\mathcal{G}^{\star}_{k} |k, x]$: the MCMC sample for $k+1$ components can be used to estimate $\Pr[\mathcal{G}^{\star}_{k} |k+1, x]$, a quantity that is related to the probability in (\ref{BF}). Let $\Pr[\mathcal{G}^{\star}_{t} |k,x]$ with $t \leq k$ be the posterior probability, conditional on $k$ components, that component $t$ is non-empty and components $t+1$ through $k$ are empty. One can show, see the Appendix, that 
\begin{equation}
	\frac{f^{\star}_{t+1}}{f^{\star}_{t}} =
	a_{t+1,t} \; \frac{\displaystyle \sum_{k = t+1}^{\kmax} 
				Pr[\mathcal{G}^{\star}_{t+1} |k,x]}
			{\displaystyle \sum_{k = t+1}^{\kmax} 
				Pr[\mathcal{G}^{\star}_{t} |k,x]}.
						\label{ratfstar}
\end{equation}
Setting $f^{\star}_{1} = f_{1}$ if available, or $f_{1}=1$ if not, the sequence $f^{\star}_{t}$ $t=1,\ldots,\kmax$ can be estimated by replacing the probabilities in (\ref{ratfstar}) with MCMC estimates. An application of (\ref{fk1}), followed by rescaling, then produces estimates of the normalized marginal likelihoods. 

Formulae (\ref{BF}) and (\ref{ratfstar}) do not assume that the prior on $(\lambda, \theta)|k$ is invariant to permutations of the components labels, only that the hyperparameters $\alpha_{j}, \phi_{j}$ are the same for all $k \geq j$. If the prior is invariant, the additional symmetry can be exploited as follows.
Let $\widetilde{\mathcal{G}}^{k}_{h} = \bigcup_{t=h}^{k} \mathcal{G}^{t}_{h}$ be the set of allocations $g$ in $\mathcal{G}_{k}$ which assign observations to exactly $h$ components. Then $\Pr[\widetilde{\mathcal{G}}^{k}_{h} |k,x]$ is the posterior probability, conditional on $k$ components, that $h \leq k$ components are non-empty. One can show, see the Appendix, that
\begin{equation}
	\frac{f^{\dagger}_{h+1}}{f^{\dagger}_{h}} = 
		(h+1) \; a_{h+1,h} \; 
		\frac{\displaystyle \sum_{k=h+1}^{\kmax} 
			\Pr[\widetilde{\mathcal{G}}^{k}_{h+1} |k,x]}
			{\displaystyle \sum_{k=h+1}^{\kmax} (k-h)
			\Pr[\widetilde{\mathcal{G}}^{k}_{h} |k,x]}.
						\label{ratfdagger}
\end{equation}
Replacing the probabilities in (\ref{ratfdagger}) with MCMC estimates and setting $f^{\dagger}_{1}$ to 1 (or $f_{1}$ if available), yields estimates of the sequence of $f^{\dagger}_{h}$'s; plugging these estimates in formula (\ref{fk3}) and rescaling produces estimates of normalized marginal likelihoods.

To illustrate the method, formula (\ref{ratfdagger}) was used to compute the marginal likelihood of $k$ components for the galaxy data. This data set consists of velocity measurements (1000 Km/sec) of 82 galaxies from the Corona Borealis region. Since its appearance in Roeder~(1990), it has been studied by several authors, see Aitkin~(2001) for an interesting comparison of likelihood and Bayesian analyses of this data set. The data was modelled as a finite mixture of univariate normals, as set out in Section~\ref{Sec-univnorm}.
The weights hyperparameters $\alpha$ were set to 1, while the other hyperparameters were $\mu=20$, $\tau=0.04$, $\gamma=2$ and $\delta=2$, their choice is discussed in Section~\ref{Sec-comparam}.
In this example I used Gibbs sampling of the allocation vectors $g$, after integrating out the weights and components parameters, see Nobile and Fearnside~(2005). However, the method applies equally well to the Gibbs sampling scheme involving both parameters and allocations, see for instance Diebolt and Robert~(1994) and Richardson and Green~(1997), as long as empty components are allowed. Each Gibbs sampler with fixed $k$ was run for 20000 sweeps, with 1000 sweeps of burn-in. The final allocation in the run with $k$ components served as the starting allocation for the run with $k+1$ components. The estimates of the marginal likelihoods normalized to sum to 1 are displayed as line-joined dots in Figure~1. For comparison, the figure also contains the estimate of the posterior of $k$ with uniform prior on $k=1,\ldots,\kmax=50$ using a different method, the allocation sampler of Nobile and Fearnside~(2005). This sampler was run for 1 million sweeps with a burn-in of 10000 sweeps and keeping only one draw every 10.
\begin{figure}[ht]
\begin{center}
\includegraphics[scale=0.65]{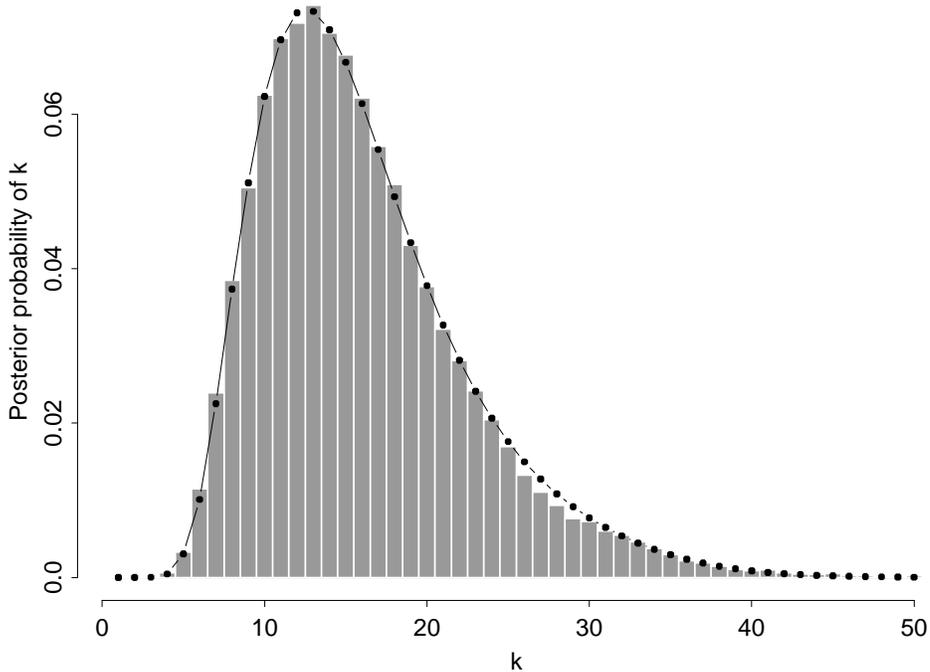}
\end{center}
\vspace{-15pt}
\caption{Two estimates of the posterior distribution of $k$ for the galaxy data, using a discrete uniform prior on $k=1,\ldots,\kmax=50$. Histogram gives the frequency of model with $k$ components using the allocation sampler of Nobile and Fearnside~(2005). Line-joined dots are the normalized marginal likelihoods using formula (\ref{ratfdagger}). See main text for hyperparameter values used.}
\label{Figure1}
\end{figure}
The agreement between the estimates from the two unrelated methods provides a welcome check on them, all the more so since visual inspection of the galaxy data suggests between three and six clusters, while the posterior of $k$ displayed in Figure~\ref{Figure1} assigns to this range of values a probability smaller than 0.02. 
If one is to believe the estimates in Figure~\ref{Figure1}, as the agreement between the two methods seems to suggest, it would seem that the posterior of $k$ has little to tell about the number of clusters in a data set.
In the next section I argue that this is not the case and that replacing the uniform prior on $k$ with a $Poi(1)$ distribution yields a posterior that is more suitable for inference about the number of actual groups in the data.

\section{The prior distribution of the number of components}
\label{Sec-priork}

\begin{figure}[ht]
\begin{center}
\includegraphics[scale=0.65]{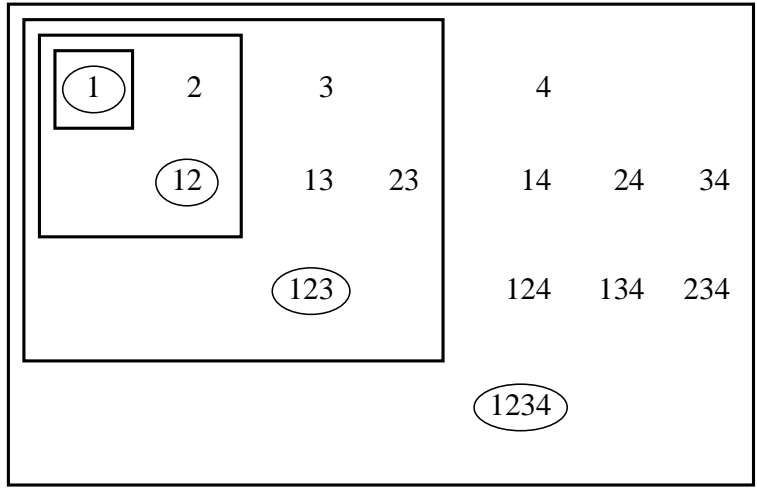}
\end{center}
\vspace{-15pt}
\caption{The nested structure of $\mathcal{G}_{k}$, $k=4$. Each set of digits denotes a set of allocation vectors with entries equal to those digits only. Nested boxes denote $\mathcal{G}_{1}$, \ldots, $\mathcal{G}_{4}$. Ellipses enclose the sets $\mathcal{G}_{h}^{h}$, $h=1,\ldots,4$. In box $\mathcal{G}_{k}$, the complement of the enclosed $\mathcal{G}_{k-1}$ box is the set $\mathcal{G}^{\star}_{k}$.}
\label{Figure2}
\end{figure}

In this section I assume that the prior on $(\lambda, \theta)|k$ is invariant to permutations of the components labels. 
Recall from Section~\ref{Sec-intro-marlik} that $f_{h}^{\dagger}$ is the part of the marginal likelihood $f_{h}$ corresponding to no empty components
\[
	f_{h}^{\dagger} = \sum_{g \in \mathcal{G}_{h}^{h}} 
				f(g|h) f(x|h,g)
\] 
and that representation (\ref{fk3}) holds:
\[
	f_{k}  =  \sum_{h=1}^{k \wedge n} 
			\binom{k}{h} a_{kh} f_{h}^{\dagger}.
\]
To have an understanding of how formula (\ref{fk3}) arises, look at Figure~\ref{Figure2} which displays the nested structure of $\mathcal{G}_{4}$, with each set of digits denoting allocation vectors with entries equal to those digits only. 
From formula (\ref{fk}), $f_{4}$ is the sum over $\mathcal{G}_{4}$ of $f(g|k=4) f(x|k=4,g)$. Formula (\ref{fk3}) gives this sum in terms of $f_{h}^{\dagger}$'s, which are sums over the sets $\mathcal{G}^{h}_{h}$, $h=1,\ldots,4$ denoted by ellipses in Figure~\ref{Figure2}. The terms $a_{kh}$ serve to rescale $f(g|h)$ to $f(g|k)$, while the combinatorial terms $\binom{k}{h}$ give the number of ``copies'' of $\mathcal{G}^{h}_{h}$ that are present in $\mathcal{G}_{k}$.

A consequence of formula (\ref{fk3}) is that the marginal likelihood of $k$ components may derive to a large extent from allocations with less than $k$ non-empty components. For instance, consider a hypothetical data set of $n=80$ observations clearly clustered in nine well separated groups, to such an extent that $f^{\dagger}_{h} / f^{\dagger}_{9}$ is nearly 0, for $h \neq 9$. Then formula (\ref{fk3}) implies that 
\begin{equation}
	\frac{f_{k}}{f_{9}} = \binom{k}{9} a_{k9}, 
	\hspace{30pt} k \geq 9.				\label{fkoverf9}
\end{equation}
With $\alpha=1$, one obtains the values reported in Table~\ref{Table1}.
\begin{table}[ht]
  \begin{center}
    \begin{tabular}{c|ccccccc}
      $k$ & 9 & 10 & 11 & 12 & 13 & 14 & 15\\
      \hline
      $f_{k} / f_{9}$ & 1. & 1.011 & 0.618 &  0.299 & 0.127 & 0.050 & 0.018
    \end{tabular}
  \end{center}
\caption{Ratio of marginal likelihood $f_{k} / f_{9}$ for a hypothetical data set with $n=80$ observations, such that $f^{\dagger}_{h} / f^{\dagger}_{9} = 0$, $h \neq 9$.}
\label{Table1}
\end{table}
With a discrete uniform prior on $k$, the posterior of $k$ gives probability less that $1/3$ to $k=9$. 
Put differently, upper bounds on the posterior of $k$ can be derived from representation (\ref{fk3}).
Table~\ref{Table2}, taken from Nobile~(2004), displays upper bounds corresponding to a discrete uniform prior and $\alpha=1$.
\begin{table}[ht]
        \begin{center}
        \begin{tabular}{|c|cccccccccc|} \hline \hline
        & \multicolumn{10}{c|}{$k$} \\
    \raisebox{8pt}{$n$} & 1 & 2 & 3 & 4 & 5 & 6 & 7 & 8 & 9 & 10 \\
    \cline{2-11}
   20 & 0.9000 & 0.7286 & 0.5299 & 0.3456 & 0.2880 & 0.2419 & 0.1954 & 
        0.1756 & 0.1505 & 0.1335 \\
   50 & 0.9600 & 0.8847 & 0.7826 & 0.6645 & 0.5414 & 0.4233 & 0.3175 & 
        0.3119 & 0.2835 & 0.2402 \\
  100 & 0.9800 & 0.9412 & 0.8858 & 0.8170 & 0.7385 & 0.6541 & 0.5677 & 
        0.4828 & 0.4023 & 0.3322 \\ 
  500 & 0.9960 & 0.9880 & 0.9762 & 0.9607 & 0.9417 & 0.9193 & 0.8938 & 
        0.8656 & 0.8350 & 0.8022 \\
        \hline \hline
        \end{tabular}
        \end{center}
        \caption{Bounds on $\pi(k|x)$ for several sample sizes $n$,
	$\pi(k)=1/k_{max}, k=1,\ldots,k_{max}=50$, $\alpha=1$.}
        \label{Table2}
\end{table}
Nobile~(2004) contains further discussion and tables for $\alpha=2$ and $\alpha=0.5$. The overall conclusion is that the bounds are weaker for larger sample sizes, smaller values of $k$ and larger values of $\alpha$. 

It is worth mentioning at this point that, as the sample size grows, the marginal likelihood of $k$ components will tend to reflect more and more only allocations with no empty components. Formally, 
\begin{equation}
	\binom{k}{h} a_{kh} \rightarrow \delta_{\{k=h\}}, 
	\hspace{25pt} n \rightarrow \infty,
							\label{asybkt}
\end{equation}
see the Appendix for a proof. Hence, from formula (\ref{fk3}), $f_{k} - f^{\dagger}_{k} \rightarrow 0$ as $ n \rightarrow \infty$.

Returning to the example of nine well separated groups, it is the combinatorial term $\binom{k}{h}$ that makes $f_{10} > f_{9}$ in Table~\ref{Table1}. 
If one were to drop the $\binom{k}{9}$ term from equation (\ref{fkoverf9}), the entries in Table~\ref{Table1} would be as in Table~\ref{Table3}.
\begin{table}[ht]
  \begin{center}
    \begin{tabular}{c|ccccccc}
      $k$ & 9 & 10 & 11 & 12 & 13 & 14 & 15\\
      \hline
      $f_{k} / f_{9}$ & 1. & 0.10112 & 0.01124 & 0.00136 & 0.00018 & 
	0.00002 & 0.00000
    \end{tabular}
  \end{center}
\caption{Ratios $f_{k} / f_{9}$ as in Table~\ref{Table1}, but dropping the term $\binom{k}{9}$ from equation (\ref{fkoverf9}).}
\label{Table3}
\end{table}
The $\binom{k}{9}$ term accounts for the fact that with $k=10$ components, there are ten possible ways of choosing nine components to have observations and one component to be empty. 
Of course, this is a consequence of the model entertained and its ability to allow for empty components, which correspond to mass on small values for some weights in the prior of $\lambda$. Nonetheless, the increasing effect on the marginal likelihoods, as $k$ grows, of the many ways in which some of $k$ components may be empty, is a rather unappealing feature of the model.
Nobile~(2004) has suggested to shift attention from the number of components to the number of non-empty components and to compute its posterior distribution.
In this paper I follow a different approach: trying to counteract the combinatorial terms in the marginal likelihoods by an appropriate choice of the prior distribution of $k$.

Multiplying equation (\ref{fk3}) by the prior distribution $\pi(k)$ and writing the result explicitly for the first few $k$, one has
\begin{eqnarray*}
	\pi(1|x) & = & A \cdot \pi(1) f^{\dagger}_{1} \\
	\pi(2|x) & = & A \left [ \pi(2) f^{\dagger}_{2}
				+ \pi(2) \binom{2}{1} a_{21} f^{\dagger}_{1}
			\right ] \\
	\pi(3|x) & = & A \left [ \pi(3) f^{\dagger}_{3}
				+ \pi(3) \binom{3}{2} a_{32} f^{\dagger}_{2}
				+ \pi(3) \binom{3}{1} a_{31} f^{\dagger}_{1}
			\right ] \\
		& \cdots &
\end{eqnarray*}
where $A$ is a normalizing constant. Although there is no prior $\pi(k)$ which exactly cancels out the binomial coefficients $\binom{k}{h}$, one can keep the contribution of $f^{\dagger}_{h}$ to $\pi(k|x)$ small, relative to its contribution to $\pi(h|x)$, by requiring that
\begin{equation}
	\pi(h) = \sup_{k > h} \left \{ \pi(k) \binom{k}{h} \right \}
	\hspace{30pt} h=1,2,\ldots\;.			\label{pihsup}
\end{equation}
It is easy to verify that a $Poi(1)$ distribution satisfies equations (\ref{pihsup}). Indeed, every prior $\pi(k)$ satisfying equations (\ref{pihsup}) is proportional to a truncated $Poi(1)$ distribution, see the Appendix. For simplicity, I will take the prior on $k$ to be $Poi(1)$.

One way to illustrate the effect of the $Poi(1)$ prior on $k$ is by recomputing the bounds on $\pi(k|x)$ with this prior; they are reported in Table~\ref{Table4}. Compared to the bounds with a discrete uniform prior in Table~\ref{Table2}, they are much weaker, especially for higher values of $k$.
\begin{table}[ht]
        \begin{center}
        \begin{tabular}{|c|cccccccccc|} \hline \hline
        & \multicolumn{10}{c|}{$k$} \\
    \raisebox{8pt}{$n$} & 1 & 2 & 3 & 4 & 5 & 6 & 7 & 8 & 9 & 10 \\
    \cline{2-11}
   20 & 0.9525 & 0.9114 & 0.8756 & 0.8441 & 0.8162 & 0.7913 & 0.7690 & 
	0.7488 & 0.7306 & 0.7140 \\
   50 & 0.9804 & 0.9619 & 0.9445 & 0.9280 & 0.9124 & 0.8976 & 0.8836 & 
        0.8703 & 0.8576 & 0.8455 \\
  100 & 0.9901 & 0.9805 & 0.9712 & 0.9621 & 0.9533 & 0.9447 & 0.9364 & 
	0.9283 & 0.9204 & 0.9128 \\ 
  500 & 0.9980 & 0.9960 & 0.9940 & 0.9921 & 0.9901 & 0.9882 & 0.9863 &
	0.9844 & 0.9825 & 0.9806 \\
        \hline \hline
        \end{tabular}
        \end{center}
        \caption{Bounds on $\pi(k|x)$ for several sample sizes $n$,
	$\pi(k) \propto 1/k!$, $\alpha=1$.}
        \label{Table4}
\end{table}

For another illustration, reconsider the example of nine well separated groups in Table~\ref{Table1}. The ratio of posterior probabilities $\pi(k|x)/\pi(9|x)$ using a $Poi(1)$ prior are given in Table~\ref{Table5}.
\begin{table}[ht]
  \begin{center}
    \begin{tabular}{c|ccccccc}
      $k$ & 9 & 10 & 11 & 12 & 13 & 14 & 15\\
      \hline
      $\pi(k|x) / \pi(9|x)$ & 1. & 0.10112 & 0.00562 & 0.00023 & 0.00001 &
	0.00000 & 0.00000
    \end{tabular}
  \end{center}
\caption{Ratio of posterior probabilities $\pi(k|x) / \pi(9|x)$ for the same hypothetical data set as in Table~\ref{Table1}, using a $Poi(1)$ prior on $k$.}
\label{Table5}
\end{table}

\begin{figure}[ht]
\begin{center}
\includegraphics[scale=0.65]{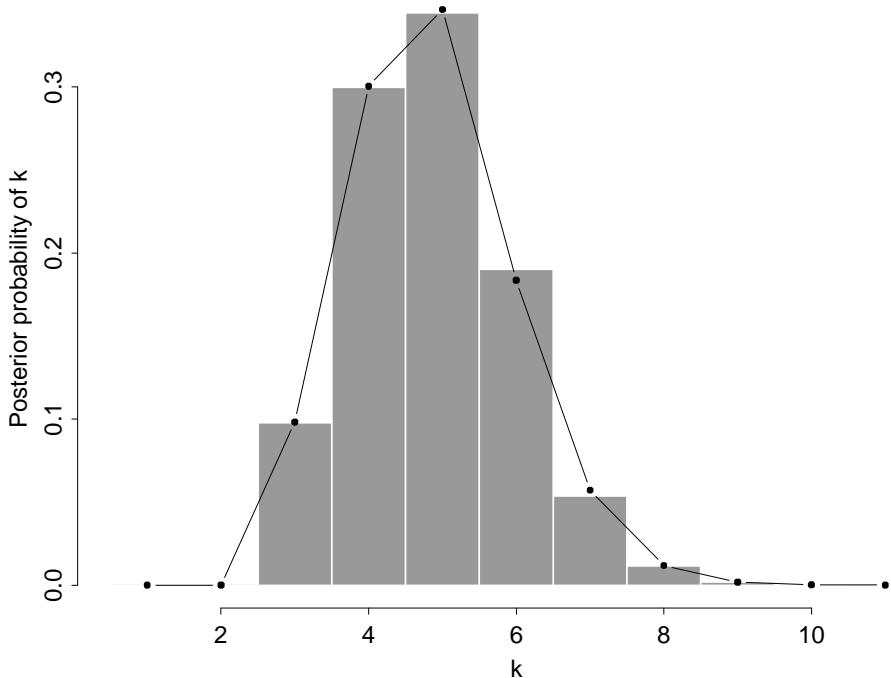}
\end{center}
\vspace{-15pt}
\caption{Two estimates of the posterior distribution of $k$ for the galaxy data, using a Poisson(1) prior. Histogram gives the frequency of model with $k$ components using the allocation sampler of Nobile and Fearnside~(2005). Solid line is the posterior of $k$ obtained from the normalized marginal likelihood using formula (\ref{ratfdagger}). See main text for hyperparameter values used.}
\label{Figure3}
\end{figure}

As a further illustration, return to the galaxy data example in Section~\ref{Sec-marlik}. Figure~\ref{Figure3} contains the posterior of $k$ computed using the same hyperparameters and methods as in Figure~\ref{Figure1}, but with a $Poi(1)$ prior on $k$, rather than discrete uniform. 
Other examples, for real and artificial data sets, of posterior distributions of $k$ based on a $Poi(1)$ prior can be found in Nobile and Fearnside~(2005).

\section{Mixtures of normals: hyperparameter selection}
\label{Sec-comparam}

This section is concerned with the choice of hyperparameters in mixtures of univarite normals, with natural conjugate priors on the means and variances. 
The method to be described can be readily adapted to the case of multivariate components, or components from other parametric families.
I continue to use the galaxy data set for illustrative purposes.
The marked sensitivity to prior specification exhibited in the analysis of this data is, in my experience, far from typical. However, it demonstrates well what difficulties may arise. Patterns of dependence of the marginal likelihood on the prior of $\theta$ are likely to be simpler in one-parameter families; see Aitkin~(2001, page 289) for a related remark.

\begin{figure}[ht]
\begin{center}
\includegraphics[scale=0.75]{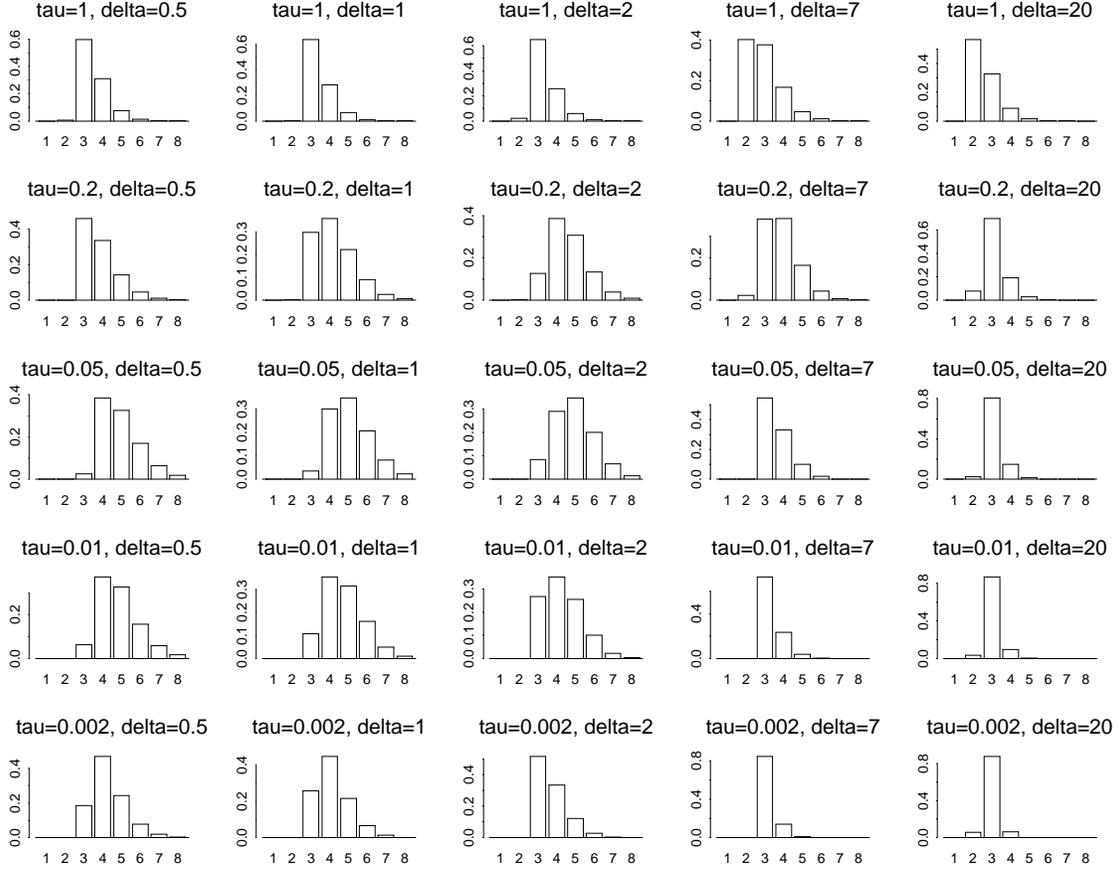}
\end{center}
\vspace{-15pt}
\caption{Posterior distribution of $k$ for the galaxy data, using a $Poi(1)$ prior, for several values of the hyperparameters $\tau$ and $\delta$.}
\label{Figure4}
\end{figure}

Figure~\ref{Figure4} displays estimates of the posterior distribution of the number of components for the galaxy data, corresponding to several values of the hyperparameters $\tau$ and $\delta$. The other hyperparameters were set to $\mu=20$ and $\gamma = 2$, as discussed in Section~\ref{Sec-univnorm}.
The prior on $k$ was $Poi(1)$ and computations were done using formula (\ref{ratfdagger}).
Although in all plots most of the mass is concentrated on values of $k$ between 2 and 8, a simple glance at the figure conveys how dependent on hyperparameter values $\pi(k|x)$ may be.
One can also see that, for given $\delta$, as $\tau$ increases at first posterior mass shifts to higher values of $k$, and then it moves back to lower values of $k$. The behaviour for $\tau$ fixed and $\delta$ increasing consists, apart for few exceptions, of a shift of probability mass from higher to lower values of $k$. Most pairs $(\tau,\delta)$ yield negligible posterior mass for $k=2$. However, some pairs in the upper right corner of the plot assign considerable mass to it. These pairs correspond to a prior distribution that makes likely high values of the variance within each normal component; in turns this makes it plausible to place in a single group the smallest and largest observations in the galaxy data, with a central group accounting for most of the other observations.

Putting a hyperprior $\pi(\tau, \delta)$ on the two hyperparameters, and sampling from the joint posterior of all the unknowns, including $\tau$ and $\delta$, did not solve the problem. 
Some experimentation with a few hyperpriors showed that $\pi(k|x)$ was to a considerable extent affected by the choice of hyperprior: the marginal posterior distributions of $\tau$ and $\delta$ had very long tails and changed markedly with $\pi(\tau, \delta)$. For this reason, I preferred to adopt an empirical Bayes stance and estimate $\tau$ and $\delta$ rather than mixing with respect to their posterior distribution.

\begin{figure}[ht]
\begin{center}
\includegraphics[scale=0.75, trim= 0 226 0 0]{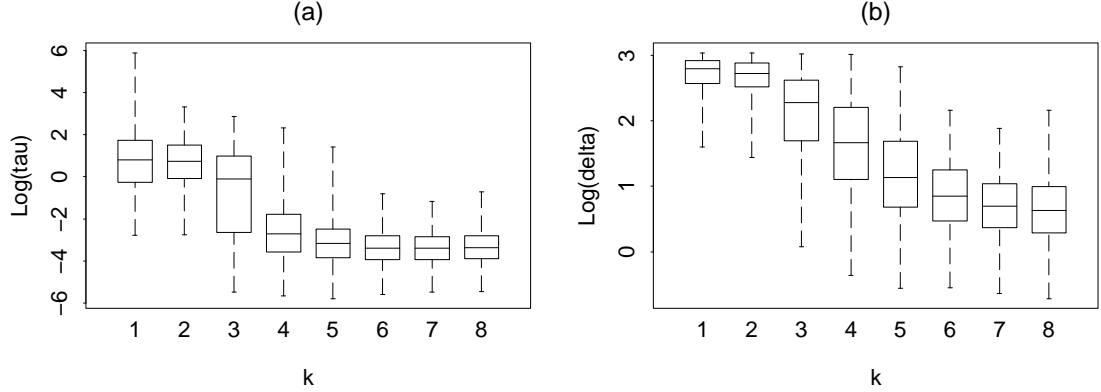}
\end{center}
\vspace{-15pt}
\caption{Boxplots of draws from the posterior distributions, conditional on $k$, of $\tau$ in panel (a) and of $\delta$ in panel (b). Whiskers are drawn at the 0.005 and 0.995 quantiles.}
\label{Figure5}
\end{figure}

I settled on independent priors: $Un(0,1)$ for $(1+\tau)^{-1}$, the prior proportion of variance within a component to the total variance, and $Un(0, \delta_{U})$ for $\delta$, where $\delta_{U} = (\gamma -1) s^{2}_{x}$ and $s^{2}_{x}$ is the sample variance. The choice of $\delta_{U}$ yields a prior expectation of the components variance equal to $s^{2}_{x} / 2$. The random walk Metropolis-Hastings algorithm was used to update $\tau$ and $\delta$ given all other variables.
Figure~\ref{Figure5} displays boxplots of the marginal posterior distributions of $\tau$ and $\delta$, on a logarithmic scale, conditional on $k$. 
Both plots display a pattern whereby a clear change of level occurs as $k$ increases. The procedure I used to estimate $\tau$ and $\delta$ consists of taking the median of the posterior draws, after discarding those corresponding to values of $k$ preceding the point where a rough level-off of the medians has occurred. The rationale is that if increasing $k$ by 1 markedly changes the posteriors of $\tau$ and $\delta$, it is because it affords a considerable reduction of the within-components variability, by replacing it with between-means variability.
The median of $\tau$ seems to level off at $k=4$. For $\delta$ the picture is less clear, but the decreases are much smaller past $k=6$. The end result are the rough estimates $\hat{\tau}=0.04$ and $\hat{\delta}=2$. These values were used in the runs reported in Sections~\ref{Sec-marlik} and \ref{Sec-priork}.
A similar procedure was used by Nobile and Fearnside~(2005), to which I refer for further examples.

The overall lesson seems to be that estimates of $\pi(k|x)$ provide only a rough, though useful, guide to the number of groups in the data and that there is really no substitute for the kind of sensitivity analysis performed in Figure~\ref{Figure4}.

\section*{Appendix}
\subsection*{A.1 \; Proof of formula (\ref{ratfstar})}

Letting $E= \mathcal{G}^{\star}_{t}$ in formula (\ref{marlikE}), with $t \leq k$, yields
\[
	f_{k} = \frac{\sum_{g \in \mathcal{G}^{\star}_{t}} 
				f(g|k) f(x|k,g)}
			{\Pr[\mathcal{G}^{\star}_{t} |k, x]}.
\]
Now from formulae (\ref{C1C2}) the numerator equals $a_{kt} f^{\star}_{t}$, so that solving for $f^{\star}_{t}$ produces
\[
	f^{\star}_{t} = \frac{\Pr[\mathcal{G}^{\star}_{t} |k, x]}
				{a_{kt}} f_{k}.
\] 
Then taking the ratio between $f^{\star}_{t+1}$ and $f^{\star}_{t}$ and rearranging terms gives
\[
	f^{\star}_{t+1} \Pr[\mathcal{G}^{\star}_{t} |k, x] =
		a_{t+1,t} \, f^{\star}_{t} \,
			\Pr[\mathcal{G}^{\star}_{t+1} |k, x]
		\hspace{30pt} t < k
\]
where one uses $a_{k,t}/a_{k,t+1} = a_{t+1,t}$, a simple consequence of (\ref{C1C2}). Summing both sides over $k$ from $t+1$ to \kmax and rearranging yields (\ref{ratfstar}).

\subsection*{A.2 \; Proof of formula (\ref{ratfdagger})}

Let $E = \widetilde{\mathcal{G}}^{k}_{h}$ in formula (\ref{marlikE}) to obtain
\[
	f_{k} = \frac{\sum_{g \in \widetilde{\mathcal{G}}^{k}_{h}} 
				f(g|k) f(x|k,g)}
			{\Pr[\widetilde{\mathcal{G}}^{k}_{h} |k, x]}
			\hspace{30pt} h \leq k.
\]
The numerator is equal to $\binom{k}{h} a_{kh} f^{\dagger}_{h}$ (see Nobile 2004, Proof of Proposition 4.3), so that rearranging one has
\[
	f^{\dagger}_{h} = \frac{ f_{k} 
				\Pr[\widetilde{\mathcal{G}}^{k}_{h} |k, x]}
				{\displaystyle \binom{k}{h} a_{kh}}
\]
Taking the ratio between $f^{\dagger}_{h+1}$ and $f^{\dagger}_{h}$ and rearranging terms produces
\[
	f^{\dagger}_{h+1} \; (k-h) \Pr[\widetilde{\mathcal{G}}^{k}_{h} |k, x] =
	f^{\dagger}_{h} \; (h+1) \; a_{h+1,h} 
		\Pr[\widetilde{\mathcal{G}}^{k}_{h+1} |k, x].
\]
Finally, sum both sides over $k=h+1, \ldots, \kmax$ and solve for $f^{\dagger}_{h+1}/f^{\dagger}_{h}$ to obtain (\ref{ratfdagger}).

\subsection*{A.3 \; Proof of formula (\ref{asybkt})}

From formula (\ref{C1C2}) and under the assumption that the prior is invariant with respect to permutations of the labels,
\[
	\binom{k}{t} a_{kt} =
		\binom{k}{t} 
		\frac{\gama{k\alpha}}{\gama{k\alpha+n}}
		 \frac{\gama{t\alpha+n}}{\gama{t\alpha}}
						\hspace{30pt} t \leq k
\]
Using formula 6.1.46 in Abramowitz and Stegun~(1964), 
$n^{(k-t)\alpha} \frac{\gama{t\alpha + n}}{\gama{k\alpha+n}} \rightarrow 1$, as $n \rightarrow \infty$. Hence
\[
	n^{(k-t)\alpha} \binom{k}{t} a_{kt} 
	\, \rightarrow \, \binom{k}{t} \frac{\gama{k\alpha}}{\gama{t\alpha}}
	\hspace{20pt} \mbox{as $n \rightarrow \infty$}.
\]	
Therefore, for $t \leq k$, $\binom{k}{t} a_{kt} \rightarrow \delta_{\{k=t\}}$ as $n \rightarrow \infty$.

\subsection*{A.4 \; Proof that every distribution satisfying equations (\ref{pihsup}) is truncated $Poi(1)$}

The proof proceeds as follows: assume that $\pi(k) = 0$ for $k$ larger than some value \kbar, use induction to derive $\pi(k)$ with $k \leq \kbar$, finally let $\kbar \rightarrow \infty$.
Suppose that, for all $k=j+1,\ldots,\kbar$, one has
\begin{equation}
	\pi(k) = \pi(\kbar) \frac{\kbar!}{k!}
							\label{Poi1pf}
\end{equation}
Then equation (\ref{Poi1pf}) also holds for $k=j$:
\begin{eqnarray*}
	\pi(j) & = & \sup_{k>j} \left \{ \pi(k) \binom{k}{j} \right \} \\
		& = & \max_{j < k \leq \kbar} 
			\left \{ \pi(\kbar) \frac{\kbar!}{k!}
				\frac{k!}{j!(k-j)!} \right \} \\
		& = &  \pi(\kbar) \frac{\kbar!}{j!}
\end{eqnarray*}
where the first line uses (\ref{pihsup}) while the second follows from (\ref{Poi1pf}) and $\pi(k) = 0$ for $k > \kbar$. Since equation (\ref{Poi1pf}) clearly holds for $k=\kbar$, an appeal to induction yields
\[
	\pi(k) = \pi(\kbar) \frac{\kbar!}{k!} \propto \frac{1}{k!}	
	\hspace{30pt} k=1,\ldots,\kbar,
\]
i.e., $Poi(1)$ restricted to $1 \leq k \leq \kbar$.
Letting $\kbar \rightarrow \infty$ yields a $Poi(1)$ distribution restricted to $k \geq 1$.

\newpage
\section*{References}

\begin{list}{}{\setlength{\itemindent}{-\leftmargin}}
\small

\item Abramowitz, M. and Stegun, I. A. (1964).
  Handbook of Mathematical Functions, Dover edition, 9th printing, New York.

\item Aitkin, M. (2001). 
	Likelihood and Bayesian analysis of mixtures. 
	{\em Statistical Modelling}, {\bf 1}, 287--304.

\item Celeux, G., Hurn, M. and Robert, C. P. (2000).
	Computational and Inferential Difficulties with Mixture
	Posterior Distributions.
	{\em Journal of the American Statistical 
	Association}, {\bf 95}, 957--970.

\item Chib, S. (1995). 
	Marginal Likelihood from the Gibbs Output. 
	{\em Journal of the American Statistical 
	Association}, {\bf 90}, 1313--1321.

\item DiCiccio, T. J., Kass, R. E., Raftery, A. and Wasserman, L. (1997).
	Computing Bayes Factors By Combining Simulation and Asymptotic 
	Approximations.
      {\em Journal of the American Statistical 
	Association}, {\bf 92}, 903--915.

\item Diebolt, J. and Robert, C. P. (1994).
      Estimation of finite mixture distributions through Bayesian sampling.
      {\em Journal of the Royal Statistical Society B},
      {\bf 56}, 363--375.

\item Fr\"{u}hwirth-Schnatter, S. (2001). 
	Markov Chain Monte Carlo Estimation of Classical and Dynamic 
	Switching and Mixture Models.  
	{\em Journal of the American Statistical Association}, 
	{\bf 96}, 194--209.

\item Gelman, A. and Meng, X.L. (1998).
	Simulating Normalizing Constants: From Importance Sampling to Bridge
	Sampling to Path Sampling.
	{\em Statistical Science}, {\bf 13}, 163--185.

\item Ishwaran, H., James, L. F. and Sun, J. (2001). 
	Bayesian Model Selection in Finite Mixtures by Marginal Density 
	Decompositions. 
	{\em Journal of the American Statistical Association}, 
	{\bf 96}, 1316--1332.

\item Nobile, A. (1994). 
        Bayesian Analysis of Finite Mixture Distributions, 
        Ph.D. dissertation, Department of Statistics, Carnegie Mellon
        Univ., Pittsburgh. Available at 
	\verb+http://www.stats.gla.ac.uk/~agostino+

\item Nobile, A. (2004).
	On the posterior distribution of the number of components 
	in a finite mixture. 
	{\em The Annals of Statistics}, {\bf 32}, 2044--2073. 
	
\item Nobile, A. and Fearnside, A. (2005). 
	Bayesian finite mixtures with an unknown number of components:
	the allocation sampler. Technical Report 05-4, 
	Department of Statistics, 
	University of Glasgow.

\item Phillips, D. B. and Smith, A. F. M. (1996).
       Bayesian model comparison via jump diffusions.
       In {\em Markov Chain Monte Carlo in Practice} (eds W. R. Gilks,
       S. Richardson and D. J. Spiegelhalter), 215--239, Chapman \& Hall.

\item Raftery, A. E. (1996).
       Hypothesis testing and model selection.
       In {\em Markov Chain Monte Carlo in Practice} (eds W. R. Gilks,
       S. Richardson and D. J. Spiegelhalter), 163--187, Chapman \& Hall.

\item Richardson, S. and Green P. J. (1997). 
       On Bayesian analysis of mixtures with an unknown number of components
       (with discussion). 
       {\em Journal of the Royal Statistical Society B},
       {\bf 59}, 731--792.

\item  Roeder, K. (1990). 
	Density estimation with confidence sets exemplified by superclusters 
	and voids in galaxies. 
	{\em Journal of the American Statistical 
	Association}, {\bf 85}, 617--624.

\item Stephens, M. (2000a). 
        Bayesian analysis of mixture models with an unknown number
        of components -- an alternative to reversible jump methods.
        {\em The Annals of Statistics}, {\bf 28}, 40--74.

\item Stephens, M. (2000b).
	Dealing with Label Switching in Mixture Models.
	{\em Journal of the Royal Statistical Society B},
        {\bf 62}, 795--809.

\end{list}

\end{document}